\documentclass[12pt]{article}
\usepackage{amssymb,amsfonts}
\usepackage[T2A]{fontenc}
\usepackage[cp866]{inputenc}
\inputencoding{cp866}
\begin{document}
\vskip1cm \centerline{\bf {SPIRAL-LOGARITHMIC STRUCTURE}}
\centerline{\bf { IN A HEISENBERG FERROMAGNET}} \vskip0.4cm
\centerline{\bf E.Sh.Gutshabash} \vskip0.3cm \centerline{\small {
Institute of Physics (Petrodvorets Branch),St.-Petersbourg State
University,}} \centerline{\small{ ul.Ulianovskaja,
1,Petrodvorets,198504 Russia}} \centerline{\small{ e-mail:
gutshab@EG2097.spb.edu}} \vskip0.3cm \centerline {\bf
{Abstract}}\vskip0.3cm \hskip1.5cm
\parbox{13cm}{\small {Spiral-logarithmic structure is suggested
as a stationary solution of a modified equation for the Heisenberg
model, and the single- and N-soliton solutions are constructed on
this base}} \vskip0.4cm PACS numbers: 75.10.-b \vskip0.5cm

Spiral (incommensurate) structure in ferromagnets have long been
studied and are well understood [1,2]. They are formed due to the
reconstruction of a Fermi surface as a result of the interaction
of magnetic moments with conduction electrons.{\footnote {The
spiral structures in non-Abelian gauge theories are discussed in
recent publication [3].}}. Mathematically, they were used in the
context of the inverse scattering problem, e.g., as boundary
conditions when solving the nonlinear (2+0)-dimensional O(3)-sigma
model (two-dimensional stationary Heisenberg ferromagnet)[4].

In this work, a special model is invoked to demonstrate
theoretically that new magnetic structures can exist which may
naturally be called spiral-logarithmic structures.

Let us consider cylindrically symmetric configurations in the
model of a "modified" Heisenberg magnet [5]:

$$
{\bf S}_t={\bf S}\times ({\bf S}_{xx}+\frac {1}{x}{\bf S}_x) \: ,\eqno(1)
$$
where ${\bf S}(x,t)=(S_1, S_2, S_3)$ is the magnetization vector ,
${\bf S}^2=1,$ and $x=\sqrt {x_1^2+x_2^2}$, where $x_1,\:x_2$ are
the Cartesian coordinates on a plane. The Cauchy problem for
Eq.(1) was solved in [5] by inverse scattering method on a trivial
background, and it was shown that any perturbation localized at
$t=0$ spreads out.

However, one can readily verify by a direct calculation that the
vector ${\bf S}={\bf S}^{(1)}$ with components

$$
S^{(1)}_1(x)=\sin(\ln x+\theta_0) \:,\:S^{(1)}_2(x)=\cos(\ln x+\theta_0)\:,
\:S^{(1)}_3(x)=0, \eqno(2)
$$
where $\theta_0\in {\mathbb R}$ is a constant and $x > 0$, is also
a (stationary) solution of Eq.(1). Now, using the fact that this
equation is quite solvable, let us construct its exact solutions
on the background of structure (2). To this end, let us represent
Eq.(1) in the form of a condition for compatibility of the
following linear matrix system:

$$
\Psi_x=U_1\Psi\Lambda,\:\;\:\Psi_t=V_1\Psi\Lambda+V_2\Psi\Lambda^2 ,\eqno(3)
$$
where $U_1=-(i/2)S,\: V_1=(1/2)S_xS+i(AS+SA),\: V_2=(i/2)S,
\:S=\sum_{i=1}^{i=3}S_i\sigma_i,\: \sigma_i$ are Pauli matrices,
$\Lambda=diag(\lambda,\:{\bar \lambda}),\:\lambda \in {\mathbb C}$
is a parameter, matrix $A=A(S)$ is determined from the condition:

$$
[A(S)S+SA(S)]_x=-\frac {1}{4ix}\bigl [S,S_x \bigr] , \eqno (4)
$$
and it is additionally assumed that the elements of the matrix
$(S-S^{(1)})$ properly decrease at $x \to {\infty}$).

Equation (1)will be solved by the Darboux matrix-transformation
method [6]. Let

$$
\tilde \Psi=\Psi-L_1\Psi\Lambda ,    \eqno(5)
$$
where $L_1=\Psi_1\Lambda_1^{-1}\Psi_1^{-1}$ and $\Psi_1$ is a
certain fixed solution of Eqs. (3) corresponding to the choice
${\bf S=S}^{(1)}$ and $ \lambda=\lambda_1$.

Checking for the covariance of system (3) about transformation (5)
yields the dressing relations ($U_1=-(i/2)S^{(1)}$):

$$
\tilde U_1=L_1U_1L_1^{-1},\;\;\;\;\tilde U_1=U_1-L_{1x}, \eqno(6)
$$
and
$$
\tilde V_2=L_1V_2L_1^{-1},\;\;\tilde V_1=V_1-L_{1t},
\;\;\tilde V_2=V_2-L_1V_1+\tilde V_1L_1 .
\eqno(7)
$$
The equivalence of all these expressions can be proved by simple
mathematics. The second relation in Eqs.(6) can be conveniently
rewritten as

$$
\tilde U_1=U_1-\Psi_1\bigl [\Psi_1^{-1}\Psi_{1x},\Lambda_1^{-1}
\bigr ]\Psi_1^{-1} .\eqno(8)
$$

From the property  $\sigma_2S\sigma_2=-\bar S$ of matrix $S$, one
has $\sigma_2U_1\sigma_2={\bar U_1}$. The matrix $\Psi_1$ then
takes the form

$$
\Psi_1=\left (\begin{array}{cc}
\varphi_1&-{\bar \chi}_1\\
\chi_1&{\bar \varphi}_1\end{array}\right) ,\eqno(9)
$$
where $\varphi_1=\varphi_1(x, t, \lambda_1)$ and $\chi_1=\chi_1(x,
t, \lambda_1)$ are complex functions to be determined.

The matrix $A$ in Eq.(4) must satisfy the condition
$Tr(AS+SA)=g(t)$ for any $S$, where $g(t)$ is the arbitrary
function (in what follows, $g(t)=0$). Furthermore, considering
that $\sigma_2SS_x\sigma_2=\overline{SS}_x$, and using (4), one
obtains ${\bar A}=\sigma_2A\sigma_2$. It can easily be seen that
Eq.(1) and the right-hand side of Eq.(4) are invariant about the
substitution $S \to S^{(f)}=S+f(x)I$, where $f(x)$ is an arbitrary
complex function and $I$ is $2\times 2$ unit matrix. By setting
$A(S)=(a_{ij}(S))$ and using condition (2), let us bring Eq. (4)
to the form $:\: diag((2ia^0_{11}(S^{(f)})f)_x,
\:\:-(2ia^0_{11}(S^{(f)})f)_x)= diag(1/(2x^2),\:\:-1/(2x^2))$,
where $a_{11}(S^{(f)})=ia^0_{11}(S^{(f)}), \:a^0_{11}={\bar
a}^0_{11}$, and the $a_{21}(S^{(f)})$ element is zero. Hence it
follows that $f(x)=i/(4x)$ and $a^0_{11}=1$; one also has
$A(S^{(f)})S^{(f)}+S^{(f)}A(S^{f})=A(S)S+SA(S)$, so that all
properties of the matrix $S$ are retained in the final formulas.

To construct the explicit solutions, note that on bare solution
(2) the system of Eqs.(3) transforms, with regard to Eq.(4), into
the following system of scalar equations $(\theta=\ln
x+\theta_0)$:

$$
\varphi_{1x}=-\frac {1}{2}\lambda_1e^{i\theta}\chi_1,
\:\:\:\:
\chi_{1x}=\frac {1}{2}\lambda_1e^{-i\theta}\varphi_1,
$$
$$
\eqno(10)
$$
$$
\varphi_{1t}=-\frac {1}{2}\lambda_1^2e^{i\theta}\chi_1,\:\:\:\:
\chi_{1t}=\frac {1}{2}\lambda_1^2e^{-i\theta}\varphi_1 .
$$
It follows from Eq.(10) that

$$
\varphi_{1t}-\lambda_1\varphi_{1x}=0\:\:,\:\:\chi_{1t}-\lambda_1\chi_{1x}=0.
\eqno(11)
$$
Therefore, $\varphi_1(x,t)=F(x+\lambda_1t)$ and $\chi_1(x,t)=
G(x+\lambda_1t)$, where functions $F$ and $G$ are constant on the
characteristic $x+\lambda_1t=const$. They can be found by solving
the system of Eqs. (10), e.g., for $t=0$. The corresponding
equation for $\varphi_1(x,0)$ is reduced to

$$
x\varphi_{1xx}-i\varphi_{1x}+\frac {1}{4}\lambda_1^2\varphi_1=0 .\eqno(12)
$$
The solution to this equation is [7]: $\varphi_1(x,0)=x^{\frac
{i}{2}+\frac {1}{2}} Z_{\frac {i}{2}+\frac {1}{2}}(-\frac
{\lambda_1x}{2})$, where $Z_{\nu}(s)=C_1I_{\nu}(s)+ C_2Y_{\nu}(s);
\:I_{\nu}$ and $Y_{\nu}$ are Bessel functions of the first and
second kind, respectively; and $C_1$ and $C_2$ are arbitrary
constants. The requirement that the solution be finite at zero
gives $C_2=0$. Then, the solution to the system of Eqs.(10) takes
the form

$$
\varphi_1(x,t)=C_1q^{\frac {i}{2}+\frac {1}{2}}I_{\frac {i}{2}+\frac{1}{2}}
(-\frac {\lambda_1}{2}q),\:\:\:
\chi_1(x,t)=-\frac {2}{\lambda_1}C_1e^{-i\theta(x,t)}\frac {\partial}
{\partial x}
[q^{\frac {i}{2}+\frac{1}{2}}I_{\frac {i}{2}+\frac{1}{2}}
(-\frac {\lambda_1}{2}q)],
\eqno(13)
$$
where $q=q(x,t)=x+\lambda_1t-x_0,\:\theta(x,t)=\ln q+\theta_0$ and
$x_0$ is the initial point.

One can now construct the simplest one-soliton solution.
Nevertheless, it is still rather cumbersome, so that it is
pertinent to use Eq.(8) and write the solution in the form
{\footnote {It is more convenient to use the matrix
$\Psi_1\sigma_3$ instead of $\Psi_1$ (it is also a solution of
system (3)).}}($S_{+}=S_1+iS_2$ and $ S_{+}[1] \equiv \tilde
S_{+}$):

$$
S_{+}[1](x,t)=ie^{-i\theta}-\frac {4\lambda_{1I}\rho_1\bigl[ (\ln
\rho_1)_x-|\rho_1|^2(\ln  {\bar \rho_1})_x\bigr]}
{|\lambda_1|^2(1+|\rho_1|^2)^2}\:,
$$
$$
{} \eqno(14)
$$
$$
S_3[1](x,t)=\frac {8\lambda_{1I}|\rho_1|^2Re (\ln \rho_1)_x}
{|\lambda_1|^2(1+|\rho_1|^2)^2}\:,
$$
where {\footnote {In this expression, as also in Eqs.(13) and
(14), the logarithms are taken on their major branches and the
appropriate cuts on the plane of parameter $\lambda \in
\mathbb{C}$ are assumed to be made to provide solution
uniqueness.}

$$\rho_1=\rho_1(x,\:t)=\chi_1/\varphi_1=-(2/\lambda_1)e^{-i\theta(x,t)}
\bigl \{{\ln\: [q^{\frac {i+2}{2}}I_{\frac {i+2}{2}}(-\lambda_1
q/2)]}\bigr \}_x.$$

This solution is nonsingular. It depends on four parameters:
$\lambda_{1R},\:\lambda_{1I},\:x_0$ and $\theta_0$, where
$\lambda_1= \lambda_{1R}+i\lambda_{1I}$, and includes a single
complex function $\rho_1(x,t)$. Note also that the spectrum of
excitations in model (1) contains satellites of the "fundamental
harmonic" with an incommensurate (and variable) "frequency". By
using the asymptotic form of the Bessel function, one can show
that the solution (14) converges to Eq.(2) at $x \to \infty$.

Let now obtain the $N$-soliton solution. For this purpose, let us
use Eq.(5). One can easily show that

$$
\Psi[N]=\Psi-Q_1\Psi\Lambda-.......-Q_N\Lambda^N,  \eqno(15)
$$
with the matrix functions $Q_1,....Q_N$ being determined from the
linear matrix system of equations

$$
\Psi_1-Q_1\Psi_1\Lambda_1^2-....-Q_N\Psi_1\Lambda_1^N=0,
$$
$$
\Psi_2-Q_1\Psi_2\Lambda_2^2-....-Q_N\Psi_2\Lambda_2^N=0,
$$
$$
{}
\eqno(16)
$$
$$
............................
$$
$$
\Psi_N-Q_1\Psi_N\Lambda_N^2-...-Q_N\Psi_N\Lambda_N^N=0,
$$
where $\Lambda_i=diag(\lambda_i,\bar \lambda_i),\:i=1...N,$ and
$\Psi_i$ are the solutions to Eq.(3) corresponding to
$\lambda=\lambda_i$. At the same time, one has from Eq. (8)

$$
U_1[N]=U_1-\sum_{i=1}^N {L_{ix}}=U_1-Q_{1x}.    \eqno(17)
$$
The expression for the $N$-soliton solution can readily be
obtained from Eqs. (16) and (17):

$$
S_3[N]=-2i(\frac {\Delta_1}{\Delta})_x ,
$$
$$
{}
\eqno(18)
$$
$$
S_{+}[N]=ie^{-i\theta}-2i(\frac {\Delta_2}{\Delta})_x ,
$$
where $S_{+}[N]=S_1[N]+iS_2[N] ,$
$$
\Delta=\left|\,\matrix{
\lambda_1\varphi_1&\lambda_1\chi_1&\lambda_1^2\varphi_1&\lambda_1^2\chi_1&
\ldots&\
\lambda_1^N\varphi_1&\lambda_1^N\chi_1\cr
-\bar \lambda_1\bar \chi_1&\bar \lambda_1\bar \varphi_1&-\bar \lambda_1^2\
\bar \chi_1&\bar \lambda_1^2\bar \varphi_1&\ldots&\
-\bar \lambda_1^N\bar \chi_1&\bar \lambda_1^N\bar \varphi_1\cr
\cdots& \cdots& \cdots& \cdots& \cdots&\cdots&\cdots&\cr
\lambda_N\varphi_N&\lambda_N\chi_N&\lambda_N^2\varphi_N&\lambda_N^2\chi_N&
\ldots&\
\lambda_N^N\varphi_N&\lambda_N^N\chi_N\cr
-\bar \lambda_N\bar \chi_N&\bar \lambda_N\bar \varphi_N&-\bar \lambda_N^2\
\bar \chi_N&\bar \lambda_N^2\bar \varphi_N&\ldots&\
-\bar \lambda_N^N\bar \chi_N&\bar \lambda_N^N\bar \varphi_N}
\right |,
$$
and $ \Delta_1$ and $\Delta_2$ are obtained by replacing the first
and second columns in the determinant $\Delta$ by
$(\varphi_1,-\chi_1,\ldots \varphi_N,-\chi_N)^T$ and $(\chi_1,\bar
\varphi_1,\ldots \chi_N\bar \varphi_N)^T$, respectively. The fact
that Eq.(18) is an $N$-soliton solution can be proved in the
standard way {it should be noted that the explicit determinant
representation is difficult to obtain by inverse scattering method
for this model, as also for the standard model of Heisenberg
magnetic, i.e., for Eq.(1)without the second term [8]}. Similar to
Eq.(14), this solution converges to the bare solution at $x \to
\infty$, while at finite $x$ values the spectrum of excitations
has on this background $N!$ coupled satellites with incommensurate
frequencies.

In conclusion, let us present a quite soluble nonlinear equation
gauge-equivalent to Eq.(1)[9]:

$$
i\psi_t+(x\psi)_{xx}+2\psi\left.\left[x\int |\psi|^2 dx \right.\right]_x=0
.\eqno(19)
$$
Due to the gauge equivalence, the analysis applied to Eq.(1) can
also be used in this case.

{\bf {Acknowledgments}}.

I am grateful to G.G.Varzugin for helpful remarks and A.B.Borisov
for the attention. This work was supported by the Russian
Foundation for Basic Research, project No.00-01-00480.

\vskip 2cm

\centerline {\bf REFERENCES}

\vskip0.5cm 1. \parbox[t]{12.7cm}
     {{\em Yu.A.Izyumov.} Neutron Diffraction by Long-Periodic Structure
     (Nauka, Moskow, 1987.)}
 \vskip0.3cm
2. \parbox[t]{12.7cm}
     {{\em L.D.Landau and E.M.Lifshits.} Course of Theoretical Physics,
     Vol.8:
     Electrodynamics of Continuous Media (Nauka, Moscow, 1982; Pergamon,
     New York, 1984.)}
\vskip0.3cm
3. \parbox[t]{12.7cm}
    {{\em R.Jackiw and So-Young Pi .} hep-th /9911072.}
 \vskip0.3cm
4. \parbox[t]{12.7cm}
     {{\em E.Sh.Gutshabash and V.D.Lipovskii.} Teor.Mat.Fiz.
     \underline {90}, 175. (1992).}
\vskip0.3cm
5. \parbox[t]{12.7cm}
     {{\em A.V.Mikhailov and A.I.Jaremchuk.} Pis'ma Zh.Eksp.Teor.Fiz.
      \underline {36}. 78. (1982).}
\vskip0.3cm
6. \parbox[t]{12.cm}
     {{ \em M.A.Salle and V.B.Matveev .} Darboux Transformation and Solitons,
      Springer-Verlag, Berlin, 1991.}
\vskip0.3cm
7. \parbox[t]{12.7cm}
     {{\em E.Kamke.} Differentialgleichungen; Losongsmethoden und
     Losungen (Geest and Portig,Leipzig,1959; Nauka, Moskow, 1961}
\vskip0.3cm
8. \parbox[t]{12.7cm}
    {{\em L.A.Takhtajan and L.D.Faddeev.} Hamiltonian Methods in the
    Theory of Solitons (Nauka, Moscow, 1986; Springer-Verlag,
    Berlin, 1987.)}
\vskip0.3cm
9.   \parbox[t]{12.7cm}
     {{\em S.P. Burtsev, V.E. Zakharov and A.V.Mikhailov.} Teor.Mat.Fiz.
      \underline{70}. 373.(1987).}

\end{document}